\documentclass[psamsfonts,twoside]{article}
\usepackage{qic-my}
\usepackage{amssymb,amsmath, amsfonts}
\usepackage[mathscr]{eucal}
\usepackage{url, hyperref}

\newcommand{\tr}{\textrm{tr}}
\newcommand{\zz}{{\mathbb Z}}
\newcommand{\x}{{\mathbf{x}}}
\newcommand{\y}{{\mathbf{y}}}
\newcommand{\cc}{{\mathbb C}}
\newtheorem{remark}{Remark}

\begin{document}
\setlength{\textheight}{8.0truein}    

\runninghead{Universal Compression of Ergodic Quantum Sources}
            {Alexei Kaltchenko and En-Hui Yang}

\normalsize\textlineskip \thispagestyle{empty}
\setcounter{page}{1}


\vspace*{0.88truein}

\alphfootnote

\fpage{1}

\vspace*{0.035truein} \centerline{\bf Universal Compression of
Ergodic Quantum Sources\footnote{This work was supported in part
by the Natural Sciences and Engineering Research Council of Canada
under Grants RGPIN203035-98 and RGPIN203035-02, by the Premier's
Research Excellence Award, by the Canada Foundation for
Innovation, by the Ontario Distinguished Research Award, and by
the Canada Research Chairs Program.}} \vspace*{0.37truein}
\centerline{\footnotesize
Alexei Kaltchenko\footnote{e-mail:
\href{mailto:akaltche@bbcr.uwaterloo.ca}{akaltche@bbcr.uwaterloo.ca}}}
\vspace*{0.015truein} \centerline{\footnotesize\it E$\&$CE
Department, University of Waterloo} \baselineskip=10pt
\centerline{\footnotesize\it Waterloo, Ontario N2L 3G1, Canada}
\vspace*{10pt} \centerline{\footnotesize En-Hui Yang}
\vspace*{0.015truein} \centerline{\footnotesize\it E$\&$CE
Department, University of Waterloo} \baselineskip=10pt
\centerline{\footnotesize\it Waterloo, Ontario N2L 3G1, Canada}
\vspace*{0.225truein} \publisher{(received date)}{(revised date)}

\vspace*{0.21truein}

\abstracts{For a real number $r>0$, let $F(r)$ be the family of
all stationary ergodic quantum sources with von Neumann entropy
rates less than $r$. We prove that, for any $r>0$, there exists a
blind, source-independent block compression scheme which
compresses every source from $F(r)$ to $r n$ qubits per input
block length~$n$ with arbitrarily high fidelity for all
large~$n$.}{As our second result, we show that the stationarity
and the ergodicity of a quantum source $\{\rho_m
\}_{m=1}^{\infty}$ are preserved by any trace-preserving
completely positive linear map of the tensor product form ${\cal
E}^{\otimes m}$, where a copy of ${\cal E}$ acts locally on each
spin lattice site. We also establish ergodicity criteria for so
called classically-correlated quantum sources.}{}

\vspace*{10pt}

 \vspace*{3pt}%
\communicate{to be filled by the Editorial}

\vspace*{1pt}\textlineskip
\section{Introduction}
The quantum ergodicity is as instrumental in studying quantum
information systems as is the classical ergodicity in studying
classical information systems. It is quite remarkable that there
are the quantum analogs of Shannon's noiseless compression
theorem\footnote{The quantum analog for i.i.d. (independently and
identically distributed) sources was first formulated and proved
in~\cite{schumacher-coding}.} and Shannon-McMillan theorem for
stationary ergodic quantum
sources\cite{Q-McMillan,bjelakovic,mosonyi}. Thus, for any such
source, one can always construct a source-dependent compression
code (scheme) which compresses the source to its von Neumann
entropy rate with arbitrarily high fidelity.

As in classical information systems\cite{davisson}, the next step
would be to see if there exists a compression scheme which does
just the same, but is source-independent, i.e. universal. A
universal scheme for the family of all i.i.d. quantum sources with
a known entropy upper bound was first introduced
in~\cite{universal-iid} and then extended to the family of all
i.i.d sources\cite{hayashi,universal-iid-unknown-entropy}. The
work\cite{hayashi} was also concerned with the scheme's optimality
and performance evaluation for every finite block length and
provided an explicit exponential bound on the encoding-decoding
error probability.

In this work we study stationary and ergodic properties of quantum
sources and present a universal compression scheme for the family
of all stationary ergodic sources with a known upper bound on
their entropy rates.

Our paper is organized as follows. In Section~\ref{Notation} we
review the mathematical formalism and notation for stationary
ergodic quantum sources. In Section~\ref{Ergodicity-Invariance} we
show that if a stationary ergodic (weakly mixing or strongly
mixing) quantum source $\{\rho_m\}_{m=1}^{\infty}$ is subjected to
a trace-preserving completely positive linear transformation (map)
of the tensor product form ${\cal E}^{\otimes m}$, where a copy of
${\cal E}$ locally acts on each spin lattice site, then all the
listed source properties are preserved. Such maps describe the
effect of a transmission via a memoryless quantum channel as well
as the effect of memoryless coding, both lossless and lossy ones.
We also establish ergodicity criteria for so called {\em
classically-correlated} quantum sources. In
Section~\ref{universal-scheme} we briefly review quantum block
compression schemes and then introduce a so-called {\em universal
projector sequence} $\left\{{p^{(n)}}\right\}_{n = 1}^\infty$ {\em
with asymptotical rate} $r>0$, where $\lim_{n \rightarrow \infty}
\frac{1}{n} \log \tr (p^{(n)}) =r$. Loosely speaking, for every
sufficiently large~$n$, the range subspace of~$p^{(n)}$ contains
the typical subspace (or high probability subspace in another
notation) for every stationary ergodic source with von Neumann
entropy rate below~$r$. This property implies the existence of a
universal compression scheme for these sources. In
Section~\ref{Universal-Construction} we prove in a constructive
way that the universal sequence of projectors does exist for any
given~$r$. The basic idea of our universal sequence construction
is as follows. We select a suitable classical
subsystem\footnote{In fact, we select an infinite family of
classical subsystems. This approach was first used to prove the
quantum analog of Shannon-McMillan theorem for completely ergodic
quantum sources\cite{mosonyi,petz} and later
extended\cite{Q-McMillan} to all ergodic sources.} of our quantum
system and restrict a given stationary ergodic quantum source to
this classical subsystem, thus obtaining a classical source. This
classical source is also stationary ergodic, and it is well-known
in classical information theory that there exist universal
compression codes for classical ergodic sources. So we select a
suitable universal classical code and then use it to construct the
universal projector sequence.
\section{Quantum Sources: Mathematical Formalism and
Notation}\label{Notation}%
Before we define a general quantum source, we give an informal,
intuitive definition of a so-called {\em classically correlated}
quantum source  as  a triple\cite{king} consisting of {\em quantum
messages}, a {\em classical probability distribution} for the
messages, and the {\em time shift}. Such a triple uniquely
determines a state of a one-dimensional quantum lattice system. If
quantum-mechanical correlation between the messages exists, one
gets the notion of a general quantum source. While any given state
corresponds to infinitely many different quantum sources, the
quantum state formalism essentially captures all the
information-theoretic properties of a corresponding quantum
source. Thus, the notion of "quantum source" is usually identified
with the notion of "state" of the corresponding lattice system and
used interchangeably.

Let~$Q$ be an infinite quantum spin lattice system over lattice
$\mathbb{Z}$ of integers.  To describe $Q$, we use the standard
mathematical formalism introduced in \cite[sec.~2.6,
defn.~2.6.3]{bratteli-1} ~\cite[sec.~6.2.1]{bratteli-2}
and~\cite[sec.~1.33 and sec.~7.1.3]{ruelle}
 and borrow notation
from~\cite{Q-McMillan} and~\cite{mosonyi}. Let ${\mathfrak A}$ be
a $C^{*}$-algebra\footnote{The algebra of all bounded linear
operators may be simply thought of as the algebra of all square
matrices with the standard matrix operations including
conjugate-transpose.} with identity of all bounded linear
operators ${\cal B}({\cal H})$ on a $d$-dimensional Hilbert space
${\cal H}$, $d < \infty$. To each $\x \in \zz$ there is associated
an algebra ${\mathfrak A}_{\mathbf{x}}$ of observables for a spin
located at site $\x$, where  ${\mathfrak A}_{\mathbf{x}}$ is
isomorphic to ${\mathfrak A}$ for every $\x$. The local
observables in any finite subset $\Lambda\subset\zz$ are those of
the finite quantum system
$$
{\mathfrak A}_{\Lambda}:= \bigotimes_{\x \in \Lambda} {\mathfrak
  A}_{\x}
$$
The quasilocal algebra ${\mathfrak A}_{\infty}$ is the operator
norm completion of the normed algebra $\bigcup\limits_{\Lambda
\subset \zz}{\mathfrak A}_\Lambda $, the union of all local
algebras ${\mathfrak A}_\Lambda$ associated with finite
$\Lambda\subset\zz$. A state of the infinite spin system is given
by a normed positive functional $$\varphi \ : \ {\mathfrak
A}_{\infty} \to \cc.$$ We define a family of states
$\{\varphi^{(\Lambda)}\}_{\Lambda \subset \zz}$, where
$\varphi^{(\Lambda)}$ denotes the restriction of the state
$\varphi$ to a finite-dimensional subalgebra ${\mathfrak
A}_\Lambda$, and assume that $\{\varphi^{(\Lambda)}\}_{\Lambda
\subset \zz}$ satisfies the so called {\em consistency}
condition{\cite{Q-McMillan, king}, that is
\begin{equation}\label{consistency}
 \varphi^{ (\Lambda)}= \varphi^{(\Lambda^{'})}
\upharpoonright{\mathfrak A}_{\Lambda} \end{equation}
 for any $\Lambda \subset
\Lambda^{'}.$ The consistent family
$\{\varphi^{(\Lambda)}\}_{\Lambda \subset \zz}$ can be thought of
as a quantum-mechanical counterpart of a consistent family of
cylinder measures. Since there is one-to-one correspondence
between the state $\varphi$ and the family
$\{\varphi^{(\Lambda)}\}_{\Lambda \subset \zz}$, any physically
realizable transformation of the infinite system~$Q$, including
coding and transmission of quantum messages, can be well
formulated using the states $\varphi^{ (\Lambda)}$ of finite
subsystems. When the subset $\Lambda \in \mathbb{Z}$ needs to be
explicitly specified, we will use the notation $\Lambda(n)$,
defined as
\begin{equation*}
\Lambda(n):= \bigl\{ x \in \mathbb{Z}: x \in \{1,\ldots,n\}
\bigr\}
\end{equation*}

Let $\gamma$ (or $\gamma^{-1}$, respectively) denote a
transformation on ${\mathfrak A}_{\infty}$ which is induced by the
right (or left, respectively) shift   on the set $\zz$. Then, for
any $l \in \mathbb{N}$, $\gamma^{l}$ (or $\gamma^{-l}$,
respectively) denotes a composition of $l$ right (or left,
respectively) shifts. Now we are equipped to define the notions of
stationarity and ergodicity of a quantum source.
\begin{definition}\label{stationarity}
A state $\varphi$ is called $N$-{\it stationary} for an integer
$N$ if $\varphi \circ {\gamma}^N=\varphi$. For $N=1$, an $N$-{\it
stationary} state is called {\it stationary}.
\end{definition}
\begin{definition}
A $N$-stationary state is called $N$-{\it ergodic} if it is an
extremal point in the set of all $N$-stationary states. For $N=1$,
$N$-ergodic state is called ergodic.
\end{definition}
The following lemma which provides a practical method of
demonstrating the ergodicity of a state is due
to~\cite[propos.~6.3.5,~Lem.~6.5.1]{ruelle}.
\begin{lemma}
The following conditions are equivalent:
\begin{enumerate}
\item[(a)] A stationary state $\varphi$ on ${\mathfrak A}_{\infty}$ is
ergodic.
\item[(b)] For all~$a, b \in {\mathfrak A}_{\infty}$, it holds
\begin{equation}\label{ergodic} \lim_{n \to \infty}
\frac{1}{n} \sum_{i=1}^n \varphi(a \ {\gamma}^i(b)) = \varphi(a) \
\varphi(b).
\end{equation}
\item[(c)] For every selfajoint~$a \in {\mathfrak A}_{\infty}$, it
holds
\begin{equation*}
\lim_{n \to \infty} \varphi\Biggl( \biggl(\frac{1}{n} \sum_{i=1}^n
{\gamma}^i(a)\biggr)^2 \Biggr) = \varphi^2(a).
\end{equation*}
\end{enumerate}
\end{lemma}
Now we state a series of definitions\cite{bratteli-1} which
provide "stronger" notions of ergodicity:
\begin{definition}
A state is called {\it completely ergodic} if it is $N$-ergodic
for every integer $N$.
\end{definition}
\begin{definition}
A stationary state $\varphi$ on ${\mathfrak A}_{\infty}$ is called
weakly mixing if
\begin{equation}\label{weakly-mixing}
\lim_{n \to \infty} \frac{1}{n} \sum_{i=1}^n \left| { \varphi(a \
{\gamma}^i(b)) -\varphi(a) \ \varphi(b) } \right| =0, \quad
\forall a, b \in {\mathfrak A}_{\infty}.
\end{equation}
\end{definition}

\begin{definition}
A stationary state $\varphi$ on ${\mathfrak A}_{\infty}$ is called
strongly mixing if
\begin{equation}\label{stronly-mixing}
\lim_{i \to \infty}   \varphi(a \ {\gamma}^i(b)) =\varphi(a) \
\varphi(b), \quad \forall a, b \in {\mathfrak A}_{\infty}.
\end{equation}
\end{definition}
It is straightforward to see that~\eqref{stronly-mixing}
$\Rightarrow$~\eqref{weakly-mixing}~$\Rightarrow$~\eqref{ergodic}.

Let $\textrm{tr}_{{\mathfrak A}_{\Lambda}}(\cdot)$ denote the
canonical trace on ${\mathfrak A}_{\Lambda}$ such that
$\textrm{tr}_{{\mathfrak A}_{\Lambda}}(e) =1$ for all
one-dimensional projections~$e$ in~${\mathfrak A}_{\Lambda}$.
Where an algebra on which the trace is defined is clear from the
context, we will omit the trace's subscript and simply write
$\textrm{tr}(\cdot)$. For each $\varphi^{(\Lambda)}$ there exists
a unique density operator $\rho_{\Lambda}\in {\mathfrak
A}_{\Lambda}$, such that
$\varphi^{(\Lambda)}(a)=\textrm{tr}(\rho_{\Lambda}a),\
a\in{\mathfrak A}_{\Lambda}$. Thus, any stationary state~$\varphi$
is uniquely defined by the consistent family of density operators
$\{\rho_{\Lambda(m)} \}_{m=1}^{\infty}$. Where no confusion
arises,  we will use the following abbreviated notation for the
rest of the paper. For all $n \in \mathbb{N}$,
\begin{alignat*}{2}
& {\mathfrak A}^{(n)} &\ := &\ {\mathfrak A}_{\Lambda(n)}\\
& \psi^{(n)} & := &\ \psi^{\Lambda(n)}\\
& \rho_n &:=  &\ \rho_{\Lambda(n)}
\end{alignat*}
It is well-known\cite{mean-entropy} in quantum mechanics that for
every stationary state~$\varphi$ the limit
\begin{equation}\label{mean-entopy-limit} s(\varphi) :=\lim_{n \to
\infty} \frac{1}{n} S\bigl(\varphi^{(n)}\bigr)
\end{equation}
exists, where $S\bigl(\varphi^{(n)}\bigr)$ is the von Neumann
entropy of the state $\varphi^{(n)}$. In quantum statistical
mechanics, the quantity $s(\varphi)$ is called the {\em mean} (von
Neumann) entropy of~$\varphi$, while in quantum information theory
it is natural to call it the {\em entropy rate} of the stationary
quantum source. It is not difficult to see that the existence of
the limit~\eqref{mean-entopy-limit} for any stationary state
implies the existence of $\lim_{n \to \infty} \frac{1}{n}
S\bigl(\psi^{(Nn)}\bigr)$  for any $N$-stationary state $\psi$ and
any fixed integer $N$. Thus, it makes possible to define a mean
entropy with respect to $N$-shift as follows
\begin{equation}\label{N-mean-entopy-limit}
s(\psi, N) :=\lim_{n \to \infty} \frac{1}{n}
S\bigl(\psi^{(Nn)}\bigr)
\end{equation}
We note that if a state is stationary, then it is also
$N$-stationary for any integer $N$. Therefore, the following
equality holds for any stationary state~$\varphi$:
\begin{equation*}
s(\varphi, N) = N s(\varphi)
\end{equation*}
\section{Invariance of Stationary and Ergodic Properties}\label{Ergodicity-Invariance}%
In this section we present a sequence of lemmas and a theorem
which help to establish the ergodicity of a state. But first we
shall reformulate the stationary ergodic properties of an infinite
spin lattice system in terms of its finite subsystems. By
rewriting the consistency condition~\eqref{consistency},
Definition~\ref{stationarity}, and the
equations~(\ref{ergodic}--\ref{stronly-mixing}) in terms of
density operators, we obtain the following three elementary
lemmas\footnote{In what follows we abusively use the same symbol
to denote both an operator (or superoperator), confined to a
lattice box $\Lambda(m)$, and its "shifted" copy, confined to a
box $\{1+j,\ldots,\ m+j\}$, where the value of integer $j$ will be
understood from the context.}.
\begin{lemma}\label{consistency-operator-form}
A family $\{\rho_{m} \}_{m=1}^{\infty}$ on ${\mathfrak
A}_{\infty}$ is consistent if and only if, for all positive
integers~$m, i<\infty$ and every $a \in {\mathfrak A}^{(m)}$, the
following holds:
\begin{equation}\label{consistency-operator-form-equation}
\tr(\rho_{m} \ a) = \tr\bigl(\rho_{m+i} \ (a \otimes I^{\otimes
i})\bigr),
\end{equation}
where $I^{\otimes i}$ stands for the $i$-fold tensor product of
the identity operators acting on respective spins.
\end{lemma}
\begin{lemma}\label{stationarity-operator-form}
A quantum source $\{\rho_{m} \}_{m=1}^{\infty}$ on ${\mathfrak
A}_{\infty}$ is stationary if and only if, for all positive
integers~$m, i<\infty$ and every $a \in {\mathfrak A}^{(m)}$, the
following equality is satisfied:
\begin{equation}
\tr(\rho_{m} \ a) = \tr\bigl(\rho_{m+i} \ (I^{\otimes i} \otimes
a)\bigr),
\end{equation}
\end{lemma}
\begin{lemma}\label{finite-ergodicity}
A stationary quantum source $\{\rho_{m} \}_{m=1}^{\infty}$ on
${\mathfrak A}_{\infty}$ is ergodic (weakly mixing or strongly
mixing, respectively) if and only if, for every positive
integer~$m <\infty$ and all $a, b \in {\mathfrak A}^{(m)}$, the
equality \eqref{ergodic1} $\bigl($\eqref{weakly-mixing1} or
\eqref{stronly-mixing1}, respectively$\bigr)$ holds:
\begin{align}
&\lim_{n \to \infty} \frac{1}{n} \sum_{i=m}^n \tr\bigl(\rho_{m+i}
\ (a \otimes I^{\otimes (i-m)} \otimes b)\bigr) = \tr(\rho_{m} a)
\ \tr(\rho_{m}
b), \label{ergodic1}  \\
&\lim_{n \to \infty} \frac{1}{n} \sum_{i=m}^n
\bigl|\tr\bigl(\rho_{m+i} \ (a \otimes I^{\otimes (i-m)} \otimes
b)\bigr) - \tr(\rho_{m} a) \ \tr(\rho_{m} b)\bigr|
=0, \label{weakly-mixing1}\\
&\lim_{i \to \infty} \tr\bigl(\rho_{m+i} \ (a \otimes I^{\otimes
(i-m)} \otimes b)\bigr) = \tr(\rho_{m}  a) \ \tr(\rho_{m}
b),\label{stronly-mixing1}
\end{align}
\end{lemma}
We now need to fix some additional notation. Let ${\cal E}$ be an
arbitrary trace-preserving completely positive linear (TPCPL)
map\cite{kraus} which takes ${\cal B}({\cal H})$ as its input.
Without loss of generality we assume that the output space for
${\cal E}$ is also ${\cal B}({\cal H})$. Next, we define a
composite map%
\[{\cal E}^{\otimes m} \ : \ {\mathfrak
A}^{(m)} \to {\mathfrak A}^{(m)}, \qquad \forall m >
0.\]%
We point out that such a tensor product map is the most
general description  of a {\em quantum memoryless
channel}\cite{barnum}.
\begin{theorem}\label{main}
If $\{\rho_{m} \}_{m=1}^{\infty}$ is a stationary and ergodic
(weakly mixing or strongly mixing, respectively) source, then so
is  the source $\left\{ { {\cal E}^{\otimes m}\bigl(\rho_{m}\bigr)
} \right\}_{m=1}^{\infty}$. The proof of this theorem is given in
the appendix~\ref{Proofs}.
\end{theorem}
\begin{remark}
This theorem can be viwed is the quantum generalization of a
well-known classical information-theoretic
result\cite[chap.~7]{berger} for memoryless channels, and we
strongly beleive that the theorem can be extended to the case of
quantum Markov channels\cite{hamada}.
\end{remark}
\begin{definition}\label{classically-correlated}
We define a classically correlated quantum source
$\{\rho_{m}^{cls} \}_{m=1}^{\infty}$ by an equation
\begin{equation}
\rho_{m}^{cls}: = \sum_{x_1 ,x_2 , \ldots ,x_m} p(x_1 ,x_2 ,
\ldots ,x_m )|x_1 \rangle \langle x_1 | \otimes |x_2 \rangle
\langle x_2 | \otimes \cdots  \otimes |x_m \rangle \langle x_m |,
\end{equation}
where $p(\cdot)$ stands for a probability distribution, and for
every $i$, $|x_i \rangle$ belongs to some fixed
linearly-independent set $S: = \left\{ {|\psi _1 \rangle ,|\psi _2
\rangle , \ldots ,|\psi _d \rangle } \right\}$ of vectors in the
Hilbert space ${\cal H}$. We recall that ${\cal H}$ is the support
space for the operators in ${\mathfrak A}$. The set $S$ is
sometimes called a {\em quantum alphabet}.
\end{definition}
\begin{corollary}\label{crllry}
If a classical probability distribution~$p(\cdot)$ in
Definition~\ref{classically-correlated} is a stationary and
ergodic (weakly mixing or strongly mixing, respectively), then so
is the quantum source $\{\rho_{m}^{cls} \}_{m=1}^{\infty}$. The
proof of this corollary is given in the appendix~\ref{Proofs}.
\end{corollary}
%
%
\section{Universal Compression with Asymptotically Perfect
Fidelity}\label{universal-scheme}%
We begin this section with the introduction of
the so-called {\em quantum block compression scheme} which
consists of a sequence $\{ {\mathcal C}^{(n)},  {\mathcal D}^{(n)}
\}_{n=1}^{\infty}$ of TPCPL compression  and decompression maps%
\begin{alignat*}{2}
&{\mathcal C}^{(n)}  \ &: & \ B({\mathcal H}^{\otimes n}) \rightarrow%
B\left({ {\mathcal H}^{n}_c }\right)\\
&{\mathcal D}^{(n)} \ &: & \ B\left({ {\mathcal H}^{n}_c }\right)
\rightarrow  B({\mathcal H}^{\otimes n}),
\end{alignat*}
where, for every integer $n>0$, ${\mathcal H}^{n}_c$ is a subspace
of a Hilbert space ${\mathcal H}^{\otimes n}$, and $B(\cdot)$
stands for the set of all linear operators on a respective Hilbert
space. For every $n$, we also define the {\em compression rate}
$r\bigl({\mathcal C}^{(n)}\bigr)$ by the equation\footnote{All
logarithms in this paper are to base~2.}
\begin{equation*}
r\bigl({\mathcal C}^{(n)}\bigr) := \frac{1}{n} \log \dim {\mathcal
H}^{n}_c.
\end{equation*}
Although a broad class of classical sources can be compressed and
decompressed without distortion i.e. with perfect fidelity,
quantum sources, with some exceptions, are not compressible
without errors\cite{incompressible}. That is, for a quantum source
$\{ \rho_m \}_{m=1}^{\infty}$, we have, in general,
\begin{equation*}
\rho_{m} \neq {\mathcal D}^{(m)}\circ{\mathcal C}^{(m)}(\rho_{m}).
\end{equation*}
However, one may still be interested in compression schemes where
the states~$\rho_{m}$ and ${\mathcal
D}^{(m)}\circ~{\mathcal~C}^{(m)}(\rho_{m})$ are sufficiently close
to each other. Such the ``closeness'' can be quantified by special
measures. In this paper, we will use the so-called {\em
entanglement fidelity}\cite{schumacher} measure which turns out to
be the ``strongest'' of all fidelity notions that are applicable
to encoding-decoding schemes. ``Strongest'' means that if the
entanglement fidelity of a compression schemes converges to unity,
then all the other applicable fidelities converge to unity,
too\cite{schumacher}. Moreover, the higher the entanglement
fidelity of a map~${\mathcal D}^{(m)}\circ{\mathcal C}^{(m)}$ is,
the better is preserved\cite{schumacher} the entanglement of the
source system with an external system. In order to define the
entanglement fidelity, we first need to introduce the {\em
fidelity of states}. The fidelity~$F(\cdot, \cdot)$ of states with
density matrices $\phi$ and $\sigma$ is defined to be
\begin{equation*}
F(\phi,\sigma) := \tr \sqrt{\phi^{\frac{1}{2}}\sigma
\phi^{\frac{1}{2}}}
\end{equation*}
Let $\phi$ be the density matrix of a state on a Hilbert space
${\mathcal H}$ which is subjected to a TPCPL map~${\mathcal E}$.
Let $|\Theta \rangle \langle \Theta | \in {\mathcal H}\otimes
\tilde{\mathcal H}$ be a purification of~$\phi$, where
$\tilde{\mathcal H}$ is a Hilbert space for the reference system
arising from purification procedure\cite{barnum0}. Then the
entanglement fidelity $F_e(\cdot, \cdot )$ is defined by
\begin{equation*}
F_e(\phi,{\mathcal E}) := F^2\bigl(|\Theta \rangle \langle \Theta
|,(\tilde{\mathcal I} \otimes {\mathcal E})(|\Theta \rangle
\langle \Theta |)\bigr),
\end{equation*}
where~$\tilde{\mathcal I}$ is the identity map (superoperator) on
the state space of the reference system. The fidelity of states
and the entanglement fidelity have many interesting
properties\cite{barnum,schumacher}, of which we just state the following%
\begin{enumerate}
\item For all $\phi$ and ${\mathcal E}$, we have the relations
\begin{equation*}
0 \leqslant F_e(\phi,{\mathcal E}) \leqslant F(\phi,{\mathcal
E}(\phi)) \leqslant 1.
\end{equation*}
\item For all $\phi$ and $\sigma$, the equality $F(\phi,\sigma) = 1$ holds if and only if $\phi =
\sigma$.
\item For all $\phi$ and ${\mathcal E}$, the equality $F_e(\phi,{\mathcal E}) = 1$ holds if and only if for all pure
states $|\psi\rangle$ lying in the support of~$\phi$,
\begin{equation*}%
{\mathcal E} (|\psi\rangle\langle\psi|) =
|\psi\rangle\langle\psi|.
\end{equation*}
\end{enumerate}
Let $\left\{ {p^{(n)} } \right\}_{n = 1}^\infty$, where $p^{(n)}
\in B({\mathcal H}^{\otimes n})$, be a sequence of orthogonal
projectors. Then we explicitly define two compression
schemes\cite{fidelity-limit,jozsa-schumacher} $\{{\mathcal
C}^{(n)}_1, {\mathcal D}^{(n)}_1 \}_{n=1}^{\infty}$ and $\{
{\mathcal C}^{(n)}_2, {\mathcal D}^{(n)}_2 \}_{n=1}^{\infty}$ as
follows
\begin{alignat*}{2}
&{\mathcal C}^{(n)}_1(\sigma)  \ &:= & \ p^{(n)}\sigma p^{(n)} + \sum_{i}{A_i\sigma A_i^{\dag}},\\
&{\mathcal C}^{(n)}_2(\sigma) \ &:= & \ \frac{p^{(n)}\sigma
p^{(n)}}{\tr\bigl(p^{(n)}\sigma p^{(n)}\bigr)},
\end{alignat*}
where $A_i$ is defined by  $A_i := |0\rangle\langle i|$, and
$\bigl\{|i\rangle \bigr\}$ is an orthonormal basis for the
orthocompliment of the subspace ${\mathcal H}^{n}_{c_1}={\mathcal
H}^{n}_{c_2}:=p^{(n)}{\mathcal H}^{\otimes n}$. The decompression
maps ${\mathcal D}^{(n)}_1$ and ${\mathcal D}^{(n)}_2$ are just
the identities on $B({\mathcal H}^{n}_{c_1})$ and $B({\mathcal
H}^{n}_{c_2})$, respectively.
\begin{definition}\label{universal-projector-sequence}
We call $\left\{ {p^{(n)} } \right\}_{n = 1}^\infty$ a {\em
universal projector sequence}  with asymptotical rate~$r
\in~\mathbb{R}$, if the following two conditions are satisfied:
\begin{enumerate}
\item  there holds the limit
\begin{equation*}
\lim_{n \rightarrow \infty} \frac{1}{n} \log \tr (p^{(n)}) =r;
\end{equation*}
\item  for {\em every} stationary ergodic source
$\{\rho_n\}_{n=1}^{\infty}$ with von Neumann entropy rate
below~$r$, the following limit also holds
\begin{equation*}
\lim_{n \rightarrow \infty} \tr (p^{(n)} \rho_n) = 1.
\end{equation*}
\end{enumerate}
\end{definition}
\begin{theorem}\label{universal-coding-theorem}
\begin{itemize}
\item[(i)] A universal projector sequence exists for any asymptotical\\ rate $r~\in~\bigl(0,\ \log\dim({\mathcal
H})\bigr]$.
\item [(ii)] Let $\left\{ {p^{(n)} } \right\}_{n = 1}^\infty$ be a universal
projector sequence with an asymptotical rate~$r$, then for {\em
every} stationary ergodic source $\{\rho_n\}_{n=1}^{\infty}$ with
von Neumann entropy rate below~$r$, the following limits hold
\begin{alignat*}{2}
\lim_{n \rightarrow \infty}  F_e( \rho_{n}, {\mathcal
D}^{(n)}_1\circ{\mathcal C}^{(n)}_1)  &= 1  & %
\lim_{n\rightarrow\infty}r\bigl({\mathcal C_1}^{(n)}\bigr) &= r\\
\lim_{n \rightarrow \infty} F_e( \rho_{n}, {\mathcal
D}^{(n)}_2\circ{\mathcal C}^{(n)}_2 )  &= 1  &\qquad
\lim_{n\rightarrow\infty}r\bigl({\mathcal C_2}^{(n)}\bigr) &= r,
\end{alignat*}
where for every $n$, the maps ${\mathcal C}^{(n)}_1$ and
${\mathcal C}^{(n)}_1$ are constructed with~$p^{(n)}$.
\end{itemize}
\end{theorem}
\begin{proof}
Part~(i) is obtained in Lemma~\ref{universal-subspace-bounds} via
explicit construction of the universal projector
sequence~\eqref{embedding}. Part~(ii) follows immediately from the
definition of a universal projector sequence and the so-called
intrinsic expression\cite{schumacher} for~$F_e(\cdot,\cdot)$ as
shown in the proof of the theorem~\cite[chap.~7,
theor.~21]{nielsen-thesis}.
\end{proof}
\begin{remark}
Theorem~\ref{universal-coding-theorem} can be viewed as a quantum
extension of universal noiseless fixed-rate
coding\cite{kieffer,ziv} of classical ergodic sources.
\end{remark}

\section{Universal Projector Sequence Construction}\label{Universal-Construction}%
We begin this section with stating the important results
from~\cite{Q-McMillan}. Since work\cite{Q-McMillan} deals with
more general, multi-dimensional quantum lattice systems,  we
reformulate the results for one-dimensional lattice, while staying
as close as possible to the original notation.
\begin{theorem}[\protect{\cite[theor.~3.1]{Q-McMillan}}]\label{decomposition}
Let $\psi$ be a stationary ergodic state. Then, for every integer
$l>1$, there exists a $k(l) \in \Lambda(l)$ and a unique convex
decomposition of $\psi$ into $l$-ergodic states~$\psi_{\x,l}$:
\begin{equation}
\psi = \frac{1}{k(l)} \sum_{\x=0}^{k(l)-1} \psi_{\x,l},
\end{equation}
where $\psi_{\x,l}$ has the following properties:
\begin{enumerate}
\item $\psi_{\x,l} = \psi_{0,l} \circ {\gamma}^{\x}$
\item $s(\psi_{\x,l}, l) = s(\psi, l)$
\end{enumerate}
\end{theorem}
\begin{lemma}[\protect{\cite[lem~3.1]{Q-McMillan}}]\label{limit-in-l}
Let $\psi$ be an ergodic state. For a real $\eta >0$, we define an
integers set $A_{l, \eta}$:
\begin{equation}\label{x-set}
A_{l, \eta} := \bigl\{ \x \in \mathbb{Z} \ : \ 0 \leqslant \x <
k(l) \ \& \ \frac{1}{l}S\bigl(\psi_{\x,l}^{(\Lambda(l))}\bigr)
\geqslant s(\psi) +\eta \bigr\}.
\end{equation}
Then, the limit
\[ \lim_{l \rightarrow \infty}\frac{|A_{l, \eta}|}{k(l)} = 0  \]
holds for every $\eta >0$, where $|A_{l, \eta}|$ denotes the
cardinality of the set~$A_{l, \eta}$.
\end{lemma}

Let $\psi$ be a stationary ergodic state on~${\mathfrak
A}^{\infty}$, and let $\{\psi_{\x,l} \}_{\x=0}^{k(l)-1}$ be the
$l$-ergodic decomposition of~$\psi$. In what follows, unless
otherwise specified, we assume that $\x$ and $\y$ are integers
from the set $\{ 0,1,\ldots,k(l)-1 \}$, where $l$ is also an
integer as defined in Theorem~\ref{decomposition}. For every $\x$,
let $\{\rho_{\x,l,n} \}_{n=1}^{\infty}$ be the family of density
operators for~$\psi_{\x,l}$, and let ${\mathfrak C}_{\x,l}$ be the
maximal abelian $C^*$-subalgebra of ${\mathfrak A}_{\Lambda(l)}$
which is generated by the spectral eigenprojections
of~$\rho_{\x,l}$. Then, the following well-known\cite{ohya}
relation holds:
\begin{equation}\label{abelian-and-non-abelian-entropy}
S(\psi_{\x,l}^{\Lambda(l)}\upharpoonright {\mathfrak C}_{\x,l}) =
S(\psi_{\x,l}^{\Lambda(l)})
\end{equation}
For a fixed $l$ and every $\x$, we define the abelian quasi-local
$C^*$-algebra ${\mathfrak C}^{\infty}_{\x,l}$, which is
constructed with the copies of~${\mathfrak C}_{\x,l}$ over the
sub-lattice $l\cdot\mathbb{ Z}$ (in the same way as ${\mathfrak
A}^{\infty}$ was constructed with~${\mathfrak A}$ over lattice
$\mathbb{Z}$) and viewed as $C^*$-subalgebra of~${\mathfrak
A}^{\infty}$, and set
\begin{alignat}{2}
&m_{\x,l} &:= &\ \psi_{\x,l} \upharpoonright {\mathfrak C}_{\x,l}^{\infty}\\
&m_{\x,l}^{(n)} &:= &\ \psi_{\x,l} \upharpoonright {\mathfrak
C}_{\x,l}^{(n)}\label{abelian-entropy-1}.
\end{alignat}
To avoid possible confusion, we emphasize that the states
$m_{\x,l}^{(n)}$ are confined to the box $\Lambda(n)$ in the
sub-lattice~$l\cdot\mathbb{ Z}$ which corresponds to the
box~$\Lambda(nl)$ in the lattice~$\mathbb{ Z}$. By
Lemma~\ref{ergodic-resriction} in the appendix, the states
$m_{\x,l}$ are stationary ergodic with respect to the
sub-lattice~$l\cdot\mathbb{ Z}$ and therefore are $l$-stationary
$l$-ergodic with respect to the lattice~$\mathbb{ Z}$. By Gelfand
isomorphism and Riesz representation, for every $\x$, quasilocal
algebra ${\mathfrak C}_{\x,l}^{\infty}$ is identified with a
measurable space which we denote by
$\bigl({\mathcal{Z}}^{\infty}_{l}, {\mathscr{P}}^{\infty}(
{\mathcal{Z}_{l}}) \bigr)$ with the following properties:
\begin{enumerate}
\item[(a)]%
the sample space ${\mathcal{Z}}^{\infty}_{l}$ is the direct
product of replicas of an abstract set $\mathcal{Z}_{l}$ with the
cardinality~$d^l$ over the the sub-lattice~$l\cdot\mathbb{ Z}$,
and ${\mathscr{P}}^{\infty}( {\mathcal{Z}})$ is the corresponding
direct product $\sigma$-field;
\item[(b)]%
there is a bijective map $f_{\x,l} : \Pi({\mathfrak
C}_{\x,l}^{\infty} ) \rightarrow  {\mathscr{P}}^{\infty}(
{\mathcal{Z}_{l}})$, where $\Pi({\mathfrak C}_{\x,l}^{\infty} )$
denotes the set of all projections in ${\mathfrak
C}_{\x,l}^{\infty}$;
\item[(c)]%
every state~$m_{\x,l}$ on ${\mathfrak C}_{\x,l}^{\infty}$
corresponds to a positive measure on
$\bigl({\mathcal{Z}}^{\infty}_{l}, {\mathscr{P}}^{\infty}(
{\mathcal{Z}_{l}}) \bigr)$ which we denote by $\mu_{\x,l}$ such
that $m_{\x,l}(p) = \mu_{\x,l}\bigl(f_{\x,l}(p)\bigr)$ for every
$p \in \Pi({\mathfrak C}_{\x,l}^{\infty}).$
\end{enumerate}
In fact, the tuple $({\mathfrak C}_{\x,l}^{\infty}, m_{\x,l})$ and
the triple $\bigl({\mathcal{Z}}^{\infty}_{l},
{\mathscr{P}}^{\infty}( {\mathcal{Z}_{l}}), \mu_{\x,l} \bigr)$ are
just two equivalent descrip\-ti\-ons\cite{ruelle} of a given
classical stochastic process (see the
appendix~\ref{commutative-algebras} for more details).
Unsurprisingly, the measure $\mu_{\x,l}$ is stationary ergodic
with respect to the sub-lattice~$l\cdot\mathbb{ Z}$, and the
following relation holds by Proposition~\ref{abelian->classic}:
\begin{equation}\label{abelian-entropy-2}
S(m^{(n)}_{\x,l})  =  H(\mu^n_{\x,l}),
\end{equation}
where $H(\mu^n_{\x,l})$ denotes the Shannon entropy of the
probability distribution on ${\mathcal{Z}}^n_{\x,l}$ defined by
measure~$\mu_{\x,l}$. It is well-known(cf.~\cite{gallager}) in
(classical) information theory that the limit $\lim_{n \to \infty}
\frac{1}{n} H(\mu^n_{\x,l})$ exists due to stationarity of
$\mu_{\x,l}$ and is called the \em Shannon entropy rate} of
measure $\mu_{\x,l}$ and denoted by $h(\mu_{\x,l})$. Shannon
entropy rate possess\cite{gallager} the following important
property
\begin{equation}\label{entropy-rate} h(\mu_{\x,l}) =\lim_{n \to
\infty} \frac{1}{n} H(\mu^n_{\x,l}) = \inf_n \frac{1}{n}
H(\mu^n_{\x,l}).
\end{equation}

For all integers $L,n>0$, let $\mathcal{U}_L$ and
$\mathcal{U}_L^n$ stand for a set of $L$ symbols (alphabet) and a
direct product of $n$ replicas of  this set, respectively. Then we
define a mapping\cite{ziv}
\begin{equation*}
C_{L,R}^n \ :  \ \mathcal{U}_L^n \rightarrow \mathcal{U}_2^{nR},
\end{equation*}
where $nR$ is an integer, and $R >0$. If $R < \log_2 L$, then some
of the elements of $\mathcal{U}_L^n$ are mapped to the same
elements of $\mathcal{U}_2^{nR}$. Let $G_{L}^n$ be a subset
of~$\mathcal{U}_L^n$ for which the mapping $C_{L,R}^n$ is
bijective. Clearly, the cardinality $|G_{L}^n|$ cannot exceed
$2^{nR}$, and we only consider mappings $C_{L,R}^n$ for which
$|G_{L}^n|$ is maximized, that is
\begin{equation*}
|G_{L}^n| = 2^{nR}.
\end{equation*}
We will denote each such set by~$G_{L,R}^n$ and call it a {\em
block code}\footnote{In some literature(cf.~\cite{ziv}) the term
``block code'' is reserved for the mapping $C_{L,R}^n$ rather than
for the set~$G_{L,R}^n$ since specifying $C_{L,R}^n$ is equivalent
to specifying $G_{L,R}^n$ up to a permutation of the alphabet.} of
{\em rate}~$R$. Now we are ready to define a so-called {\em
universal sequence of block codes} on a measurable space
$\bigl(\mathcal{U}_L^{\infty}, \mathcal{F}^{\infty}(\mathcal{U}_L)
\bigr)$, where $\mathcal{F}^{\infty}(\mathcal{U}_L)$ is the usual
product
$\sigma$-field of subsets of~$\mathcal{U}_L^{\infty}$.%
\begin{definition}
A sequence of block codes $\{G_{L,R}^n \}_{n=1}^{\infty}$ on
$\bigl(\mathcal{U}_L^{\infty}, \mathcal{F}^{\infty}(\mathcal{U}_L)
\bigr)$ is called {\em universal} if for every stationary ergodic
measure $\mu$ on $\bigl(\mathcal{U}_L^{\infty},
\mathcal{F}^{\infty}(\mathcal{U}_L) \bigr)$ with $h(\mu) < R$
there holds the limit%
\begin{equation}\label{limit-of-probab}
\lim_{n \to \infty} \mu\bigl( G^n_{L,R}\bigr) =1.
\end{equation}
\end{definition}
The existence of a universal sequence for any real $R>0$ was
shown\cite{kieffer,ziv} in the framework of the universal
(classical) compression of stationary ergodic sources.
It is not difficult to see that, for any $R >0$, a universal
sequence $\{G_{L,R}^n \}_{n=1}^{\infty}$ can be
constructed\cite{MPM} in such a way that for any integer $i >0$,
the subsequence $\{G_{L,R}^{i\cdot j} \}_{j=1}^{\infty}$ gives
rise to another universal sequence $\{G_{L^i,iR}^j
\}_{j=1}^{\infty}$. More specifically, if we partition every
sequence in $G_{L,R}^{i\cdot j}$ into non-overlapping blocks of
length~$i$, and view it as the sequence of the supersymbols, then
we get exactly the set~$G_{L^i,iR}^j$. From now on, we will be
only considering universal sequences with this property.

Now, for all $l$ and $R >0$, let
$\left\{{\Omega_{l,R}^{(n)}}\right\}_{n=1}^{\infty}$ be a
universal sequence on $\bigl({\mathcal{Z}}^{\infty}_{l},
{\mathscr{P}}^{\infty}( {\mathcal{Z}_{l}}) \bigr)$, and, for every
$\x$, let~$p_{\x,l,R}^{(n)}$ be a projector in~${\mathfrak
C}_{\x,l}^{(n)}$ that corresponds to the set~$\Omega_{l,R}^{(n)}$,
that is
\begin{equation}\label{projector-set-correspondence}
p_{\x,l, R}^{(n)} := f_{\x,l}^{-1}\Bigl(\Omega_{l,R}^{(n)}\Bigr)
\end{equation}
Let~$\tilde \psi$ be an arbitrary stationary ergodic state
on~${\mathfrak A}^{\infty}$, and we convert all the notation we
introduce in connection with the sate $\psi$ to the notation for
$\tilde \psi$ by adding the symbol \~{}. We know that given any
two (faithful) states on ${\mathfrak A}^{(l)}$, for any~$l$, the
eigenbasis of the density operator of one state can be obtained
from the other state's eigenbasis by applying some unitary
operator in ${\mathfrak A}^{(l)}$. Therefore,  for every pair
$\x$~and~$\y$, there exists a unitary operator~$U_{l} \in
{\mathfrak A}^{(l)}$ which satisfies the equality
\begin{equation}\label{rotated}
\tilde p_{\y,l, R}^{(n)} = U_{l}^{\otimes n} p_{\x,l, R }^{(n)} \
U_{l}^{\dag \otimes n},
\end{equation}
We define an auxiliary  projector $w_{l,R}^{(ln)} \in {\mathfrak
A}^{(ln)}$
\begin{equation}\label{rotated-projector-2}
w_{l,R}^{(ln)} : = \bigvee_{U_{l} \in {\mathfrak A}^{(l)}}
U_{l}^{\otimes n} p_{0,l,R}^{(n)} \ U_{l}^{\dag \otimes n},
\end{equation}
where $p_{0,l,R}^{(n)} := \left. {p_{\x,l,R}^{(n)}}
\right|_{\x=0}$. Then, for every $\x$ and all real $R>0$, we have
\begin{equation}\label{rotated-projector-1}
\tilde p_{\x, R}^{(n)} \leqslant w_{l,R}^{(ln)}.
\end{equation}
Moreover, for all integer $i,j >0$ and real $R>0$, we have the inequality%
\begin{equation}\label{univ-projector1}
w_{l,R}^{(l\cdot ij)} \leqslant w_{il, iR}^{(il\cdot j)}
\end{equation}
due to the special relationship between the universal codes
$G_{L,R}^{i\cdot j}$ and $G_{L^i,iR}^j$, which we discussed
earlier. Finally, we construct, for any real~$r>0$, a projector
sequence $\{ q_r^{(m)} \}_{m=1}^{\infty}$ as follows. For every
integer $m>0$, let $i_m$ be the integer-valued function of~$m$
which is defined by the inequality
\begin{equation}
2^{i_m}d^{3\cdot2^{i_m}} \leqslant m <
2^{i_m+1}d^{3\cdot2^{i_m+1}},
\end{equation}
and we also define integer-valued functions $l_m, n_m$ and
real-valued function $R_m$ via equalities
\begin{alignat*}{2}
&l_m &:= &\ 2^{i_m}, \\
&n_m &:= &\ \left\lfloor{ \frac{m}{l_m}  }\right\rfloor, \\
&R_m \ &:= &\ l_m\cdot r.
\end{alignat*}
Then $q_r^{(m)}$ is given by the expression%
\begin{equation}\label{embedding}
q_r^{(m)} := \left\{ {\begin{array}{*{20}l}
   {w_{l_m,R_m}^{(l_mn_m)}} & {\text{if} \ m= 2^{i_m}d^{3\cdot2^{i_m}}},\\
   {w_{l_m,R_m}^{(l_mn_m)} \otimes I^{\otimes(m-l_mn_m)}} & {\text{otherwise}}.\\
 \end{array} } \right.
\end{equation}
Thus, projectors $q^{(m)}_r$ do not depend on either ~$\psi$
or~$\tilde \psi$ or any other state(s).
\begin{lemma}\label{universal-subspace-bounds}
For any real $0< r\leqslant\log d$ and stationary ergodic source
$\psi$ with $s(\psi)<r$, the following two limits hold:
\begin{itemize}
\item[(i)]$\lim_{m \rightarrow \infty} \psi^{(m)}(q^{(m)}_r) = 1$;
\item[(ii)]$\lim_{m \rightarrow \infty} \frac{1}{m} \log \tr
(q^{(m)}_r)=r$.
\end{itemize}
%
\end{lemma}
\begin{proof}%
For all integer $m \geqslant \tilde m >0$, we have the following
sequence of relations
\begin{alignat}{2}%
&\psi^{(m)} \left( { q^{(m)}_r } \right) \ &\overset{1)}{=}
& \ \psi^{(l_mn_m)}\left({w_{l_m,R_m}^{(l_mn_m)}}\right)%
\overset{2)}{\geqslant} \psi^{(l_mn_m)}\left( { w_{l_{\tilde
m},R_{\tilde m}}^{(l_mn_m)} }\right)%
\overset{3)}{=}  \frac{1}{k(l_{\tilde m})}
\sum_{\x=0}^{k(l_{\tilde m})-1}
\psi_{\x,l_{\tilde m}}^{(l_mn_m)}\left({w_{l_{\tilde m},R_{\tilde m}}^{(l_mn_m)}}\right)\notag\\
&  &\geqslant & \frac{1}{k(l_{\tilde m})} \sum_{\x \in
A_{l_{\tilde m}, \eta}^c}
\psi_{\x,l_{\tilde m}}^{(l_mn_m)}\left( { w_{l_{\tilde m},R_{\tilde m}}^{(l_mn_m)}}\right)%
\geqslant \frac{|A_{l_{\tilde m}, \eta}^c|}{k(l_{\tilde m})}
\min_{\x \in A_{l_{\tilde m}, \eta}^c} \psi_{\x,l_{\tilde
m}}^{(l_mn_m)}\left( { w_{l_{\tilde m},R_{\tilde
m}}^{(l_mn_m)}}\right)\notag\\
&  &\overset{4)}{\geqslant} & \frac{|A_{l_{\tilde m},
\eta}^c|}{k(l_{\tilde m})} \min_{\x \in A_{l_{\tilde m}, \eta}^c}
\psi_{\x,l_{\tilde m}}^{(l_mn_m)}\left( {p_{\x,l_{\tilde
m},R_{\tilde
m}}^{(l_mn_m/ l_{\tilde m})}}\right)%
= \frac{|A_{l_{\tilde m}, \eta}^c|}{k(l_{\tilde m})} \min_{\x \in
A_{l_{\tilde m}, \eta}^c} \mu_{\x,l_{\tilde m}} \biggl(
\Omega_{l_{\tilde m},R_{\tilde m}}^{(l_mn_m/ l_{\tilde m})}
\biggr)\label{prob-bound}
\end{alignat}
where~$1)$ is due to~\eqref{embedding}, $2)$ is due
to~\eqref{univ-projector1}, $3)$ is due
to~Theorem~\ref{decomposition}, $4)$ is due
to~\eqref{rotated-projector-1}, and $\Omega_{l_{\tilde
m},R_{\tilde m}}^{(j)} \equiv f_{\x,l_{\tilde m}}\left( {
p_{\x,l_{\tilde m},R_{\tilde m}}^{(j)}} \right)$ for any integer
$j>0$ by definition~\eqref{projector-set-correspondence}.

\noindent Now we want to show that for every $\x \in A_{l_{\tilde
m},\eta}^c$, where $\eta := r- s(\psi)$, the inequality
\begin{equation}\label{entropy-bound-2}
h(\mu_{\x,l_{\tilde m}}) < R_{\tilde m}
\end{equation}
holds. First, we upper-bound $h(\mu_{\x,l})$,  for all integer $l$
and $\x$, as follows
\begin{alignat}{2}\label{entropy-bound-1}
h(\mu_{\x,l}) \overset{a)}{\leqslant} H(\mu_{\x,l}^{(1)})
\overset{b)}{=} S(\psi_{\x,l}^{(l)}\upharpoonright {\mathfrak
C}_{\x,l})\overset{c)}{=} S(\psi_{\x,l}^{(l)}),
\end{alignat}
where $a)$ is due to~\eqref{entropy-rate},  $b)$ is due
to~\eqref{abelian-entropy-1},~\eqref{abelian-entropy-2}, and $c)$
is due to~\eqref{abelian-and-non-abelian-entropy}. Then
\eqref{entropy-bound-1} and \eqref{x-set}
imply~\eqref{entropy-bound-2}.

Since $l_m$ is a non-decreasing function of~$m$, and there holds
the limit
\begin{equation}\label{limit-for-l-m}
\lim_{m \rightarrow \infty} l_m = \infty,
\end{equation}
Lemma~\ref{limit-in-l} implies the existence of the limit
\begin{equation*}
\lim_{m \rightarrow \infty}\frac{|A_{l_m, \eta}^c|}{k(l_m)} = 1.
\end{equation*}
That is, for any $\epsilon >0$, there exists an integer $\tilde
m_{\epsilon, \eta} >0$ which satisfies the inequality
\begin{equation}\label{epsilon1probab-bound}
\frac{|A_{l_{\tilde m_{\epsilon, \eta}}, \eta}^c|}{k(l_{\tilde
m_{\epsilon, \eta}})} > 1 - \epsilon.
\end{equation}
On the other hand, for every integer $m > \tilde m_{\epsilon,
\eta}$, the expression $l_mn_m/ l_{\tilde m}$ is a non-decreasing
integer-valued function of~$m$, and there holds the limit
\begin{equation*}
\lim_{m \rightarrow \infty} l_mn_m = \infty.
\end{equation*}
Then by~\eqref{limit-of-probab} there exists an integer
$M_{\epsilon, \eta} >\tilde m_{\epsilon, \eta}$ such that for
every integer $m > M_{\epsilon, \eta}$ and every $\x \in
A_{l_{\tilde m_{\epsilon, \eta}},\eta}^c$, there holds the
inequality
\begin{equation}\label{epsilon2probab-bound}
\mu_{\x,l_{\tilde m_{\epsilon, \eta}}} \biggl( \Omega_{l_{\tilde
m_{\epsilon, \eta}},R_{\tilde m_{\epsilon, \eta}}}^{(l_mn_m/
l_{\tilde m_{\epsilon, \eta}})} \biggr) > 1 - \epsilon.
\end{equation}
Thus, combining~\eqref{prob-bound}, \eqref{epsilon1probab-bound},
and~\eqref{epsilon2probab-bound}, we obtain the first part of the
lemma.

%
%
To prove the second part of the lemma, we will make use of the
simple upper bound\cite{universal-iid} on the dimensionality of a
so-called {\em symmetrical subspace} of a linear space. We define
a space
\[ SYM\bigl({\mathfrak A}^{(ln)}\bigr):=
span \bigl\{A^{\otimes n} : A \in {\mathfrak A}^{(l)} \bigr\},\]
which is the symmetrical subspace of~${\mathfrak A}^{(ln)}$ over
sub-lattice box~$l\cdot\Lambda(n)$. Then the dimensionality
of~$SYM\bigl({\mathfrak A}^{(ln)}\bigr)$ is
upper-bounded\cite{universal-iid} by $(n+1)^{d^{2l}}$. Thus, for
all integer $m>0$ and real~$r>0$, we have
\begin{alignat*}{2}
&\tr \left( { q^{(m)}_r} \right) &= & \tr \left( {
w_{l_m,R_m}^{(l_mn_m)}} \right) \cdot \tr \left( { I^{\otimes (m-l_mn_m)}}\right)%
\leqslant SYM\bigl({\mathfrak A}^{(l_mn_m)}\bigr)\cdot \tr\left( {p_{0,l_m,R_m}^{(n_m)}}\right)\cdot d^{l_m}\\
& &\leqslant & (n_m+1)^{d^{2l_m}} \cdot \bigl|
\Omega_{0,l_m,R_m}^{(n_m)} \bigr| \cdot d^{l_m} =
(n_m+1)^{d^{2l_m}} \cdot 2^{n_mR_m} \cdot d^{l_m},
\end{alignat*}
and
\begin{alignat}{2}
&\frac{1}{m}\log\tr\left({q^{(m)}_r}\right)%
\ &\leqslant& \ \frac{1}{l_mn_m}\log\tr\left({q^{(m)}_r}\right)%
\overset{1)}{\leqslant}
\frac{d^{2l_m}\log(d^{3l_m}+1)}{l_md^{3l_m}} +r +\frac{\log
d}{d^{3l_m}}\label{dimensions-upper-bound},
\end{alignat}
where $1)$ is due to the fact that inequality $n_m \geqslant
d^{3l_m}$ holds for all integers $m>0$. On the other hand,
\begin{alignat}{2}
&\frac{1}{m}\log\tr\left({q^{(m)}_r}\right)%
\ &\overset{1)}{\geqslant} \ \frac{n_mR_m + (m-l_mn_m)\log d}{m}%
\overset{2)}{\geqslant} r\label{dimensions-lower-bound},
\end{alignat}
where $1)$ is due \eqref{rotated-projector-1} and $2)$ is due to
the relation $rl_m \equiv R_m \leqslant l_m \log d$ which holds
for all~$m$. Combining~\eqref{limit-for-l-m},
\eqref{dimensions-upper-bound}, and
\eqref{dimensions-lower-bound}, we obtain the second part of the
lemma.
\end{proof}

\section{Conclusion}
We prove that, for any real number $r>0$, there exists a sequence
$\left\{{p^{(n)}}\right\}_{n = 1}^\infty$ of orthogonal projectors
such that for any stationary ergodic source with von Neumann
entropy rate below $r$ and all sufficiently large~$n$, the range
subspace of~$p^{(n)}$ approximately contains the source's typical
subspace. Thus, we can compress the source by projecting it into
the range subspace. Since $\left\{{p^{(n)}}\right\}_{n =
1}^\infty$ does not depend on the source, we obtain a universal
compression scheme for the family of all stationary ergodic
sources with the entropy rates less than $r$. This extends the
result\cite{universal-iid} obtained by Jozsa~et~al.  for
independently and identically distributed quantum sources.

We also show invariance of stationary and ergodic properties under
completely positive linear transformations that describe the
effect of a transmission via a quantum memoryless channel. As the
corrolarly of our invariance result, we establish ergodicity
criteria for classically-correlated quantum sources. This can be
viewd as a step towards the studies on how the properties of a
quantum source are changed after transmission through a quantum
channel, and which subclasses of stationary ergodic quantum
sources are invariant under certain transformations.

\myappendix{ States on Quasilocal Commutative
$C^*$-algebras}\label{commutative-algebras}%
Let $\mathfrak{B}$ be an arbitrary commutative $k$-dimensional
$C^*$-subalgebra of~${\cal B}({\cal H})$, and let
$\mathfrak{B}_{\infty}$ be a quasilocal algebra
$\mathfrak{B}_{\infty}$ over lattice $\mathbb{Z}$ with local
algebras $\mathfrak{B}_{\x}$ isomorphic to $\mathfrak{B}$ for
every $\x \in \mathbb{Z}$, i.e., $\mathfrak{B}_{\infty}$ is
constructed in the same way as is ${\mathfrak A}_{\infty}$ in
Section~\ref{Notation}. Then, for any $\Lambda \subset
\mathbb{Z}$, every minimal projector in $\mathfrak{B}_{\Lambda}$
is necessarily one-dimensional, and the density operator for every
pure state $\varphi^{(\Lambda)}$ on $\mathfrak{B}_{\Lambda}$ is
exactly a one-dimensional projector. Let $\bigl\{ |z_i\rangle
\langle z_i| \bigr\}_{i=1}^{k}$ be a collection of the density
operators for all the distinct pure states on $\mathfrak{B}$. We
then define an abstract set $\mathcal{Z}:=\{ z_i \}_{i=1}^{k}$,
where every element $z_i$ symbolically corresponds to the
operator~$|z_i\rangle \langle z_i| $, and $z_i \neq z_j$ for all
$i \neq j$. For every finite lattice subset $\Lambda \in
\mathbb{Z}$, we define the Cartesian product
$$
{\mathcal{Z}}^{\Lambda}:=  \underset{\x \in
\Lambda}{\pmb{\times}}{\mathcal{Z}}_{\x},
$$
i.e., the elements $\omega$ of ${\mathcal{Z}}^{\Lambda (n)}$ have
the form $\omega = \omega_1\ldots\omega_{n}$, $\omega_i \in
\mathcal{Z}$. It is easy to see that, for every $\Lambda \in
\mathbb{Z}$, the set ${\mathcal{Z}}^{\Lambda}$ and the set of all
one-dimensional projectors in $\mathfrak{B}_{\Lambda}$ are in
one-to-one correspondence: $\omega \longleftrightarrow
|\omega\rangle \langle \omega|$. Consequently, there is one-to-one
correspondence between the set of all projectors in
$\mathfrak{B}_{\Lambda}$ and ${\mathscr{P}}^{\Lambda}(
{\mathcal{Z}})$, the Cartesian product of the power sets of
${\mathcal{Z}}$. In particular,  every projector $p \in
\mathfrak{B}_{\Lambda}$ corresponds to a set $\bigl\{ \omega :
\omega \in {\mathcal{Z}}^{\Lambda}, |\omega\rangle \langle \omega|
\leqslant p \bigr\}$. We note that, equipped with the product of
the discrete topologies of the sets ${\mathcal{Z}}_{\x}$,
${\mathcal{Z}}^{\Lambda}$ is a compact space, and the pair
$\bigl({\mathcal{Z}}^{\Lambda}, {\mathscr{P}}^{\Lambda}(
{\mathcal{Z}}) \bigr)$ defines a measurable space. Thus, by
Gelfand-Naimark theorem\cite[chap.~11]{rudin2} and Riesz
representation theorem\cite[sec.~2.14]{rudin1}, for any pure or
mixed state $\varphi^{(\Lambda)}$ on $\mathfrak{B}_{\Lambda}$,
there exists a unique positive measure on
$\bigl({\mathcal{Z}}^{\Lambda}, {\mathscr{P}}^{\Lambda}(
{\mathcal{Z}}) \bigr)$, denoted by $\mu_{\Lambda}$, such that the
following equality holds for any projector $p \in
\mathfrak{B}_{\Lambda}$:
\begin{equation}\label{state<->measure}
\varphi^{(\Lambda)} (p)= \sum_{|\omega\rangle \langle \omega|
\leqslant p} \mu_{\Lambda} (\omega)
\end{equation}
Combining~\eqref{state<->measure} and
\eqref{consistency-operator-form-equation} and setting $a:=
|\omega_1\ldots\omega_m\rangle \langle \omega_m\ldots\omega_1|$ in
the latter, we obtain, for any $m, i \in \mathbb{N}$ and any
$\omega_{1}\ldots\omega_{m} \in {\mathcal{Z}}^{\Lambda (m)}$,
\begin{equation}\label{classical-consistency-cond}
\mu_{\Lambda (m)} (\omega_{1}\ldots\omega_{m})=
\sum_{\omega_{m+1}\ldots\omega_{m+i}} \mu_{\Lambda (m+i)}
(\omega_{1}\ldots\omega_{m} \omega_{m+1}\ldots\omega_{m+i})
\end{equation}
The equality~\eqref{classical-consistency-cond} is called the
(classical) {\em consistency} condition. Thus,
$\{\mu_{\Lambda}\}_{\Lambda \subset \zz}$ is a consistent family
of probability measures, and $\mu_{\Lambda}$ extends to a
probability measure  on $\bigl({\mathcal{Z}}^{\infty},
{\mathscr{P}}^{\infty}( {\mathcal{Z}}) \bigr)$ by the Kolmogorov
extension theorem\cite{kolmogorov}. The extended measure is
denoted by~$\mu$.
\begin{propos}\label{abelian->classic}
If a state $\varphi$ on $\mathfrak{B}_{\infty}$ is stationary and
ergodic (weakly mixing or strongly mixing, respectively), then so
is the corresponding measure $\mu$ on
$\bigl({\mathcal{Z}}^{\infty}, {\mathscr{P}}^{\infty}(
{\mathcal{Z}}) \bigr)$, and the following entropy relations hold:
\begin{alignat}{2}
& S(\varphi^{(n)}) & = &\ H(\mu^n)\\
& s(\varphi) & =  &\ h(\mu)
\end{alignat}
where $H(\mu^n)$ and $h(\mu)$ denote the Shannon entropy of the
probability distribution on ${\mathcal{Z}}^n$ defined by
measure~$\mu$ and the Shannon entropy rate of $\mu$, respectively.
The converse is also true.
\end{propos}
\begin{proof}
The result follows immediately from
Lemma~\ref{stationarity-operator-form},
Lemma~\ref{finite-ergodicity}, and  the
equality~\eqref{state<->measure}.
\end{proof}

\myappendix{ Conditional expectation}%
Let $\tilde {\mathfrak A}$ be a $C^*$-subalgebra of~${\mathfrak
A}$, and let $E : {\mathfrak A} \rightarrow \tilde{\mathfrak A}$
be a linear mapping which sends the density of every state
$\varphi$ on ${\mathfrak A}$ to the density of the state~$\varphi
\upharpoonright \tilde{\mathfrak A}$. Such a mapping is usually
called a {\em conditional expectation} and has the following
properties\cite[propos.~1.12]{ohya}:
\begin{enumerate}
\item[(a)] if $a \in {\mathfrak A}$ is positive operator,
then so is $E (a) \in \tilde{\mathfrak A}$;
\item[(b)] $E (b) = b$ for every $b \in \tilde{\mathfrak A}$;
\item[(c)] $E (ab) = E (a) b$ for every~$a \in {\mathfrak
A}$ and~$b\in \tilde{\mathfrak A}$;
\item[(d)] for every~$a \in {\mathfrak A}$, it holds \[\tr_{{\mathfrak A}}(a) = \frac{\tr_{{\mathfrak A}}(I)}{\tr_{\tilde{\mathfrak
A}}(I)} \tr_{\tilde{\mathfrak A}}\bigl(E (a)\bigr),  \] where $I$
stands for identity operator.
\end{enumerate}
\begin{lemma}\label{ergodic-resriction}
Let ${\mathfrak A}^{\infty}$ be the quasi-local $C^*$-algebra,
which is constructed with the copies of the finite-dimensional
$C^*$-algebra ${\mathfrak A}$ over the lattice $\mathbb{Z}$ as
described in Section~\ref{Notation}. Let ${\mathfrak C}$ be a
maximal abelian $C^*$-subalgebra of ${\mathfrak A}$, and let
${\mathfrak C}^{\infty} \subset {\mathfrak A}^{\infty}$ be the
abelian quasi-local $C^*$-algebra which is constructed with the
copies of ${\mathfrak C}$ over the lattice $\mathbb{Z}$. Then, for
every stationary ergodic state~$\varphi$ on~${\mathfrak
A}^{\infty}$, the state $\varphi \upharpoonright {\mathfrak
C}^{\infty}$ is also stationary ergodic.
\end{lemma}
\begin{proof}%
For any integer $m>1$, let $E_m : {\mathfrak A}^{(m)} \rightarrow
{\mathfrak C}^{(m)}$ be the conditional expectation mapping which
sends the density of $\varphi^{(m)}$ to the density of
$\varphi^{(m)} \upharpoonright {\mathfrak C}^{(m)}$, and let
$\{\rho_m \}_{m=1}^{\infty}$ be the family of density operators
for $\varphi$. Since~${\mathfrak C}^{(m)}$ is a maximal abelian
subalgebra of~${\mathfrak A}^{(m)}$, we have $\tr_{{\mathfrak
A}^{(m)}}(I) = \tr_{{\mathfrak C}^{(m)}}(I)$. Then, the following
equalities hold by the properties of conditional expectation for
all positive integers $m < i < \infty$ and all $a, b \in
{\mathfrak C}^{(m)}$:
\begin{align*}
&\tr_{{\mathfrak A}^{(m+i)}} \Bigl(  \rho_{m+i} \bigl(a\otimes I^{\otimes(i-m)}\otimes b\bigr) \Bigr)%
= \tr_{{\mathfrak C}^{(m+i)}}\bigg( E_{m+i}\Bigl(  \rho_{m+i} \bigl(a\otimes I^{\otimes(i-m)}\otimes b\bigr) \Bigr) \bigg)\\%
= \  &\tr_{{\mathfrak C}^{(m+i)}}\Big( E_{m+i}(  \rho_{m+i}) \bigl(a\otimes I^{\otimes(i-m)}\otimes b\bigr)  \Big),  \\
&\tr_{{\mathfrak A}^{(m)}} (\rho_m a) = \tr_{{\mathfrak C}^{(m)}}
\bigl(E_m(\rho_m a)\bigr) = \tr_{{\mathfrak C}^{(m)}}
\bigl(E_m(\rho_m) a\bigr),\\
&\tr_{{\mathfrak A}^{(m)}} (\rho_m b) = \tr_{{\mathfrak C}^{(m)}}
\bigl(E_m(\rho_m b)\bigr) = \tr_{{\mathfrak C}^{(m)}}
\bigl(E_m(\rho_m) b\bigr).
\end{align*}
Thus, the family $\{ E_m(\rho_m) \}_{m=1}^{\infty}$ is consistent,
stationary, and ergodic by the
lemmas~\ref{consistency-operator-form},
\ref{stationarity-operator-form}, and~\ref{finite-ergodicity}.
\end{proof}
\myappendix{ Proofs}\label{Proofs}%
\noindent\textbf{Proof of Theorem
\ref{main}:}\\
For any TPCPL map there exists a so-called "operator-sum
representation"\cite{barnum},\cite{kraus}. Then, an $m$-fold
tensor product map ${\cal E}^{\otimes m}$ has the following
representation:
\begin{equation}\label{operation-representation1}
{\cal E}^{\otimes m}\bigl(\rho_{m}\bigr)
=\sum_{j_1,j_2,\ldots,j_m} \bigl(A_{j_1} \otimes A_{j_2}
\otimes\cdots\otimes A_{j_m} \bigr) \rho_{[1,m]} \bigl(A_{j_1}
\otimes A_{j_2} \otimes\cdots\otimes A_{j_m}
\bigr)^{\dag}\end{equation}%
with
\begin{equation}\label{operation-representation2}
\sum_{i} A_i^{\dag} A_i = I, \quad A_{i}, I \in {\mathfrak A},
\end{equation}%
where $I$ stands for identity operator.\\
Due to~\eqref{operation-representation1} and
\eqref{operation-representation2}, the following three equalities
hold for all positive integers~$m <i <\infty$ and all $a, b \in
{\mathfrak A}^{(m)}$
\begin{equation}\label{set}
\begin{gathered}
\tr\bigl({\cal E}^{\otimes (m+i)}(\rho_{m+i}) \ (a \otimes
I^{\otimes (i-m)} \otimes b)\bigr) = \tr\bigl( \rho_{m+i} \ (
\tilde a \otimes I^{\otimes (i-m)} \otimes
\tilde b )\bigr),\hfill \\
\tr({\cal E}^{\otimes m}(\rho_{m})  a) =
\tr(\rho_{m}  \tilde a),\hfill \\
\tr({\cal E}^{\otimes m}(\rho_{\Lambda(m)})  b) =
\tr(\rho_{m}  \tilde b),\hfill \\
\end{gathered}
\end{equation} where $a, b \in {\mathfrak A}^{(m)}$ and $\tilde a$
and $\tilde b$ are defined as follows:
\begin{gather*}
\tilde a := \sum_{j_1,j_2,\ldots,j_m} \bigl(A_{j_1} \otimes
A_{j_2} \otimes\cdots\otimes A_{j_m} \bigr)^{\dag} a \bigl(A_{j_1}
\otimes A_{j_2} \otimes\cdots\otimes A_{j_m}
\bigr), \\
\tilde b :=  \sum_{j_1,j_2,\ldots,j_m} \bigl(A_{j_1} \otimes
A_{j_2} \otimes\cdots\otimes A_{j_m} \bigr)^{\dag} b \bigl(A_{j_1}
\otimes A_{j_2} \otimes\cdots\otimes A_{j_m}
\bigr). \\
\end{gather*}
Combining~\eqref{set} with Lemma~\ref{finite-ergodicity}, we
obtain the ergodicity (weakly mixing or strongly mixing,
respectively) of $\left\{ { {\cal E}^{\otimes
m}\bigl(\rho_{m}\bigr) } \right\}_{m=1}^{\infty}$. In a similar
manner, the application of Lemma~\ref{consistency-operator-form}
establishes  the consistency of $\left\{ { {\cal E}^{\otimes
m}\bigl(\rho_{m}\bigr) } \right\}_{m=1}^{\infty}$, and the
application of Lemma~\ref{stationarity-operator-form} establishes
the stationarity of $\left\{ { {\cal E}^{\otimes
m}\bigl(\rho_{m}\bigr) } \right\}_{m=1}^{\infty}$.
$\square$
%

\noindent\textbf{Proof of Corollary
\ref{crllry}:}\\
Let $S_{\bot}:= \left\{ {|e _1 \rangle ,|e _2 \rangle , \ldots ,|e
_d \rangle } \right\}$ be any orthonormal basis in~${\cal H}$, and
let $\{\tilde\rho_{m}^{cls} \}_{m=1}^{\infty}$ be the source with
alphabet $S_{\bot}$ and distribution $p(\cdot)$. For
$i=1,\ldots,d$, we define a set $\{A_i\}$ of linear operators as
follows
\begin{equation}
A_i:=|\psi _i \rangle \langle e_i|.
\end{equation}
Then, set $\{A_i\}$ satisfies \eqref{operation-representation2},
and we define a TPCPL map ${\cal E}^{\otimes m}$ as
in~\eqref{operation-representation1}. Consequently, we have
$\bigl(\rho_{m}^{cls}\bigr) = {\cal E}^{\otimes m}\bigl(\tilde
\rho_{m}^{cls}\bigr)$. Thus, to complete the proof, we need to
show that $\{\tilde \rho_{m}^{cls} \}_{m=1}^{\infty}$ on
${\mathfrak A}_{\infty}$ is ergodic (weakly mixing or strongly
mixing, respectively). Let ${\mathfrak C}$ be a subalgebra of
${\mathfrak A}$ spanned by the set $\{ |e_i\rangle \langle e_i| \
: \ |e_i\rangle \in S_{\bot} \}$. We extend ${\mathfrak C}$ to a
quasilocal algebra ${\mathfrak C}_{\infty} \subset {\mathfrak
A}_{\infty}$ over lattice $\zz$ in the same way we did for
${\mathfrak A}_{\infty}$. The algebra ${\mathfrak C}_{\infty}$ is
abelian due to the orthogonality of the set $S_{\bot}$.
For any integer $m>1$, let $E_m : {\mathfrak A}^{(m)} \rightarrow
{\mathfrak C}^{(m)}$ denote the conditional expectation.
Since~${\mathfrak C}^{(m)}$ is a maximal abelian subalgebra
of~${\mathfrak A}^{(m)}$, we have $\tr_{{\mathfrak A}^{(m)}}(I) =
\tr_{{\mathfrak C}^{(m}}(I)$. Moreover, by our construction,
$\tilde \rho_{m}^{cls}$ is an element of algebra ${\mathfrak
C}^{(m)} \subset {\mathfrak A}^{(m)}$ for every $m$. Then, the
following equalities hold by the properties of conditional
expectation for all positive integers $m < i < \infty$ and all $a,
b \in {\mathfrak A}^{(m)}$:
\begin{align*}
&\tr_{{\mathfrak A}^{(m+i)}} \Bigl(\tilde \rho^{cls}_{m+i} \bigl(a\otimes I^{\otimes(i-m)}\otimes b\bigr) \Bigr)%
= \tr_{{\mathfrak C}^{(m+i)}}\bigg( E_{m+i}\Bigl(  \tilde \rho^{cls}_{m+i} \bigl(a\otimes I^{\otimes(i-m)}\otimes b\bigr) \Bigr) \bigg)\\%
= \  &\tr_{{\mathfrak C}^{(m+i)}}\Big( \tilde \rho^{cls}_{m+i} E_{m+i}\bigl(a\otimes I^{\otimes(i-m)}\otimes b\bigr) \Big),  \\
&\tr_{{\mathfrak A}^{(m)}} (\tilde \rho^{cls}_m a) =
\tr_{{\mathfrak C}^{(m)}} \bigl(E_m(\tilde \rho^{cls}_m a)\bigr) =
\tr_{{\mathfrak C}^{(m)}}
\bigl(\tilde \rho^{cls}_m E_m(a)\bigr),\\
&\tr_{{\mathfrak A}^{(m)}} (\tilde \rho^{cls}_m b) =
\tr_{{\mathfrak C}^{(m)}} \bigl(E_m(\tilde \rho^{cls}_m b)\bigr) =
\tr_{{\mathfrak C}^{(m)}} \bigl(\tilde \rho^{cls}_m E_m(b)\bigr).
\end{align*}
Thus, if $\{\tilde \rho_{m}^{cls} \}_{m=1}^{\infty}$ is
consistent, stationary, and ergodic (weakly mixing or strongly
mixing, respectively) on ${\mathfrak C}_{\infty}$, then it also
holds on ${\mathfrak A}_{\infty}$ by the
lemmas~\ref{consistency-operator-form},
\ref{stationarity-operator-form}, and~\ref{finite-ergodicity}.
Finally, we note that since ${\mathfrak C}_{\infty}$ is abelian,
$\{\tilde \rho_{m}^{cls} \}_{m=1}^{\infty}$ on ${\mathfrak
C}_{\infty}$ is ergodic (weakly mixing or strongly mixing,
respectively) if and only if so is~$p(\cdot)$ by
Proposition~\ref{abelian->classic} from the
appendix~\ref{commutative-algebras}. $\square$
%

\nonumsection{References}%

\end{document}